\documentclass[]{aastex631}

\usepackage{graphicx}	
\usepackage{subfigure} 
\usepackage{amsmath}	
\usepackage{balance}

\begin{document}

\title[The rate and luminosity function of FRBs]{The formation rate and luminosity function of fast radio bursts}

\author{J. H. Chen}
\affiliation{School of Astronomy and Space Science, Nanjing University, Nanjing 210093, China}

\author{X. D. Jia}
\affiliation{School of Astronomy and Space Science, Nanjing University, Nanjing 210093, China}

\author{X. F. Dong}
\affiliation{School of Astronomy and Space Science, Nanjing University, Nanjing 210093, China}

\author{F. Y. Wang}
\affiliation{School of Astronomy and Space Science, Nanjing University, Nanjing 210093, China}
\affiliation{Key Laboratory of Modern Astronomy and Astrophysics (Nanjing University), Ministry of Education, Nanjing 210093, China}

\correspondingauthor{F. Y. Wang (fayinwang@nju.edu.cn)}



\begin{abstract}

Fast radio bursts (FRBs) are millisecond-duration flashes with unknown origins. Its formation rate is crucial for unveiling physical origins. However, the luminosity and formation rate are degenerated when directly fitting the redshift distribution of FRBs. In contrast to previous forward-fitting methods, we use the Lynden-Bell's $c^{-}$ method to derive luminosity function and formation rate of FRBs without any assumptions. Using the non-repeating FRBs from the first CHIME/FRB catalog, we find a relatively strong luminosity evolution, and luminosity function can be fitted by a broken power-law model with a break at $1.33\times10^{41}\ \mathrm{erg}\ \mathrm{s}^{-1}$. The formation rate declines rapidly as $(1+z)^{-4.9\pm0.3}$ with a local rate $1.13\times10^4\ \mathrm{Gpc}^{-3}\ \mathrm{yr}^{-1}$. This monotonic decrease is similar to the rate of short gamma-ray bursts. After comparing it with star formation rate and stellar mass density, we conclude that the old populations including neutron stars and black holes, are closely related to the origins of FRBs. Monte Carlo simulations are used to test our results. The distributions of mock sample are consistent with the observational data.
\end{abstract}



\section{Introduction}
Fast radio bursts (FRBs) are a type of transient radio pulses that last milliseconds in the universe flashing randomly. They were first discovered in 2007 \citep{2007Sci...318..777L}. 
Later, several other similar radio pulses were discovered \citep{2013Sci...341...53T}, which made FRBs attract great attention in the field of astronomy. At present, more than 700 FRBs have been detected between 110 MHz and 8 GHz, which can be divided into repeaters and non-repeaters in general, and a lot of models for FRBs have been proposed \citep{2021SCPMA..6449501X,2023RvMP...95c5005Z}. However, the physical origin of them is unclear. 

The discovery of the Galactic FRB 20200428 from SGR 1935+2154 suggests that at least some FRBs are produced by magnetars from deaths of massive stars \citep{2020Natur.587...59B,2020Natur.587...54C}. Therefore, the viewpoint that FRBs follow the history of cosmic star formation has been widely adopted. 
However, a repeating source FRB 20200120E was discovered in an M81 globular cluster \citep{2021ApJ...910L..18B,2022Natur.602..585K,2022NatAs...6..393N}, which indicates that some of FRBs are related to ancient star populations and not directly associated with young populations.  
The rapid growth of detected FRB events enables us to utilize the statistical characteristics to study the physical origins of them.

The relation between the formation rate of FRBs and star formation rate (SFR) will shed light on their progenitors. \cite{2019MNRAS.487.3672Z} studied the energy function and formation rate of FRBs by including the metallicity effect for Parkes sample and Australian Square Kilometer Array Pathfinder (ASKAP) sample. They got a moderate time delay (about 3–5 Gyr) relative to cosmic SFR. \cite{2024arXiv240506281W} also found that FRB rate evolves with SFR with a short delay time. 
The number density of FRB sources can be compared with the density of possible ancestors to constrain FRB sources \citep{Cao2018,2019NatAs...3..928R,2020MNRAS.494.2886H,2020MNRAS.494..665L}. In addition, the redshift evolution of the luminosity or energy function of FRBs is also one of the useful tools for constraining the progenitors of FRBs \citep[]{2020MNRAS.498.3927H,2021MNRAS.501.5319A,2022MNRAS.510L..18J,2022MNRAS.511.1961H}. 
Using the first CHIME/FRB catalog \citep{2021ApJS..257...59C}, \cite{2022ApJ...924L..14Z} excluded the hypothesis that all FRBs track the SFR. \cite{2022MNRAS.511.1961H} found that the event rate of non-repeating FRBs is likely controlled by the old populations, such as neutron stars and black holes. 
However, all above works use the forward-fitting method, in which the observed data, such as the fluence, and redshift distributions, are fit to predictions of models with assumptions for luminosity functions and rate of FRBs. Here, we will determine luminosity function and rate of FRBs from the data directly using the Lynden-Bell's $c^{-}$ method.


Lynden-Bell's $c^{-}$ method is a non-parametric data processing approach without any assumptions \citep[]{1971MNRAS.155...95L}. 
It can deal with the flux-limit sample, which has been widely applied in  quasars \citep[]{1971MNRAS.155...95L,1992ApJ...399..345E}, long gamma-ray bursts (GRBs) \citep[]{2002ApJ...574..554L,2012MNRAS.423.2627W,2015ApJ...806...44P,2015ApJS..218...13Y,2016A&A...587A..40P,2019MNRAS.488.5823L,2022MNRAS.513.1078D} , short GRBs \citep[]{Yonetoku2014,2018ApJ...852....1Z} and galaxies \citep[]{1978AJ.....83.1549K,1986MNRAS.221..233P}. \cite{2019JHEAp..23....1D} has used this method to derive energy function and formation rate of FRBs observed by Parkes and ASKAP. But the sample is relatively small. 

In this paper, we derive the luminosity function and rate of FRBs using Lynden-Bell's $c^{-}$ method with the CHIME/FRB catalog. This paper is organized as follows. In Section \ref{FRB sample}, we describe the FRB sample. In Section \ref{Lynden-bell's}, Lynden-Bell's $c^{-}$ method is introduced. The luminosity function and formation rate are derived in Section \ref{Luminosity function and formation rate of FRBs}. Monte Carlo simulation is presented in Section \ref{Testing with Monte Carlo simulation}. Conclusions and discussion are shown in Section \ref{Conclusions and discussions}. Here, we adopt the flat $\mathrm{\Lambda CDM}$ model with $\Omega_{\rm m}=0.315$ and $H_{0}= 67.4\ \mathrm{km~s^{- 1}}\mathrm{Mpc^{- 1}}$.

\section{FRB sample}
\label{FRB sample}
The Canadian Hydrogen Intensity Mapping Experiment (CHIME) can detect FRBs in the 400-800 MHz frequency band \citep[]{2018ApJ...863...48C}. CHIME/FRB collaboration released a catalog of 536 FRBs between 400 and 800 MHz from July 25, 2018 to July 1, 2019, referred to as "Catalog 1" \citep[]{2021ApJS..257...59C}. With Catalog 1, we now have the ability to conduct statistical research with uniform selection effects.

In order to accurately reflect the intrinsic population of FRBs, we select apparently non-repeating FRBs. Considering that currently non-repeating sources and repeating ones may have different origins, we choose 474 non-repeating FRBs. Some FRBs with small extra-galactic dispersion measures (DM) are discarded. Because their redshifts are hard to determined from DM-z relation. Finally there are totally 439 FRBs in our sample. \cite{2023ChPhC..47h5105T} derived the pseudo redshifts for FRBs in Catalog 1 using the probability distributions of dispersion measures from cosmological simulations \citep{Zhang2020,Zhang2021}. Their results are used.

\begin{figure}
\centering
    \includegraphics[width=0.5\textwidth]{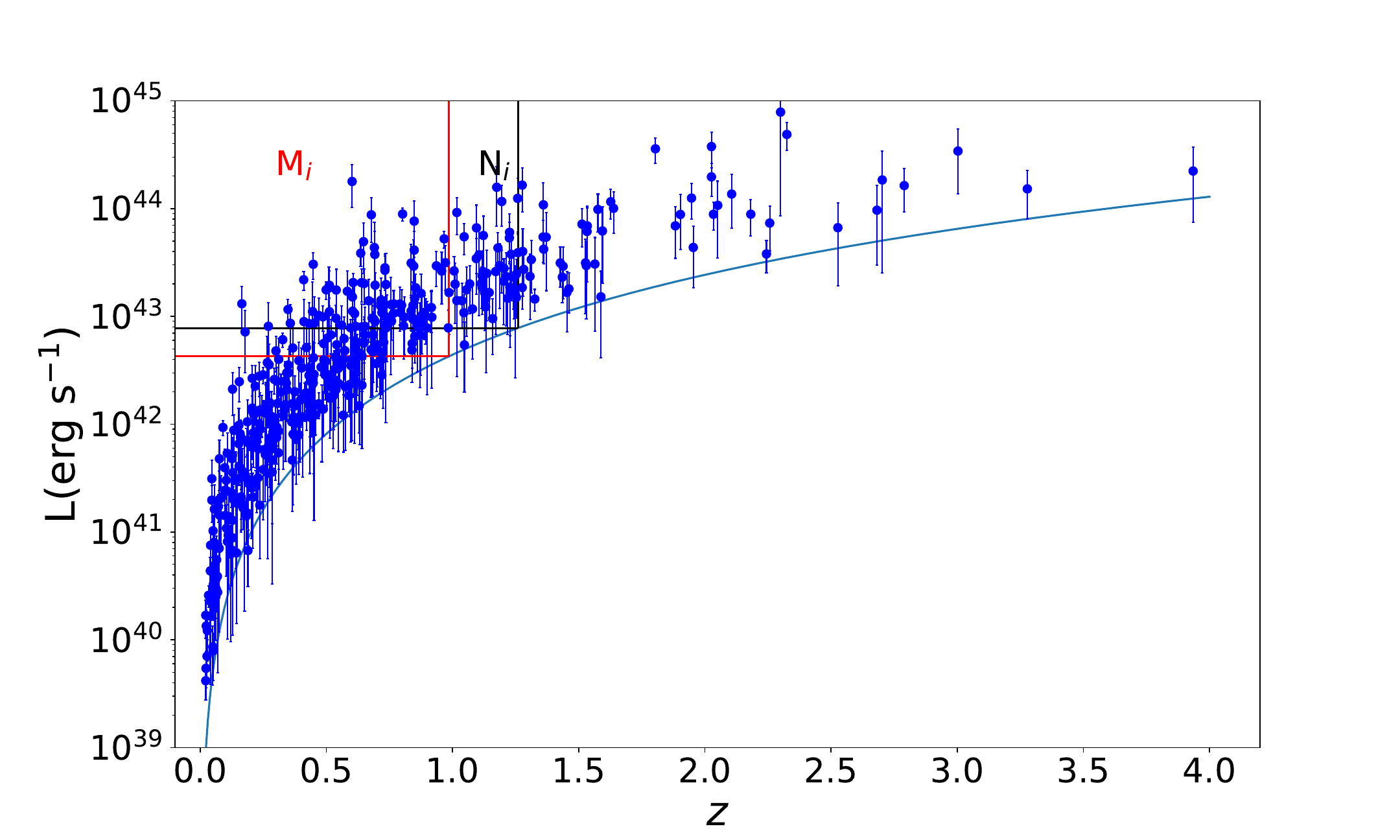}
    \caption{Luminosity-redshift distribution of FRBs. The blue dots represent FRBs and the blue solid line is the flux limit of 0.2 Jy. The error bars are derived by flux errors in CHIME Catalog 1. $M_{i}$ is shown in the red rectangle and $N_{i}$ is shown in the black rectangle.}
    \label{L-z}
\end{figure}

The luminosity can be calculated from
\begin{equation}
L=4\pi d_{L}^{2}(z)F\Delta\nu,
\end{equation}
where
\begin{equation}
d_L(z)=\frac {c(1+z)}{H_0}\int_0^z\frac{dz}{\sqrt{1-\Omega_\mathrm{m}+\Omega_\mathrm{m}(1+z)^3}}
\end{equation}
is the luminosity distance at redshift $z$. In this equation, $F$ represents the peak flux observed between a certain frequency range and $\Delta\nu=400\ \mathrm{MHz}$ is the bandwidth of CHIME. 
We adopt the flux limit as 0.2 $\mathrm{Jy}$ \citep{2018ApJ...863...48C,2022ApJ...924L..14Z}. The luminosity distribution of FRBs is shown in Figure ~\ref{L-z}.

\begin{figure}
\centering
	\includegraphics[width=0.5\textwidth]{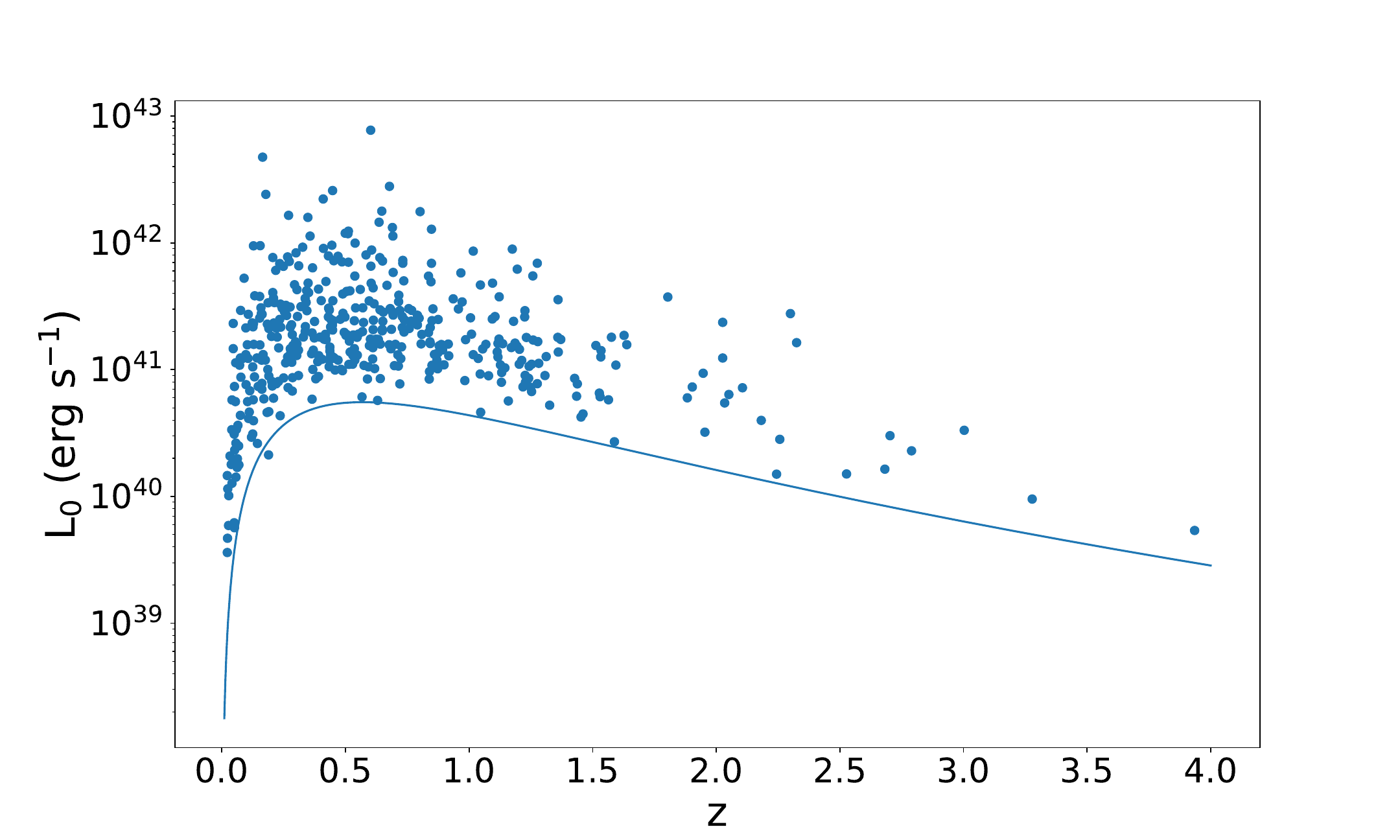}
    \caption{Non-evolving luminosity $L_0=L/(1+z)^{6.66}$ of 439 FRBs. The blue dots represent FRBs and the blue solid line represents the observational limit considering the impact of $k$.}
    \label{L0-z.pdf}
\end{figure}

\section{Lynden-Bell's $c^{-}$ method}
\label{Lynden-bell's}
Lynden-Bell's $c^{-}$ method is an effective method for determining the luminosity and redshift distribution of astronomical objects using truncated data samples. This method can break the degeneracy between the luminosity function and formation rate. If the luminosity and redshift of FRBs are independent, then we can rewrite $\Psi(L,z)$ as $\Psi(L,z)=\psi(L)\phi(z)$, where $\psi(L)$ represents the cumulative luminosity function and $\phi(z)$ is the cumulative redshift distribution of FRBs. However, the luminosity and redshift of FRBs are not independent. We should remove the effect of the luminosity evolution $g(z)=(1+z)^{k}$ and the transformed luminosity $L_{0}=L/g(z)$ is independent with redshift. As a result, we can obtain $\Psi(L,z)=\psi_{z}(L)\phi(z)=\psi(L_{0})\phi(z)$. The luminosity function of a FRB at redshift $z$ is $\psi_{z}(L)=\psi(L/g(z))$.

The first step is to obtain the form of the evolution function $g(z)$. In order to  calculate the value of $k$, we introduce a "non parametric" test method \citep[]{1992ApJ...399..345E}, which has been widely used in previous works \citep[]{Lloyd-Ronning_2002,2015ApJS..218...13Y}. For each point $(L_{i},z_{i})$ in the statistic, we can define the associated set $J_i$ as \citep[]{1992ApJ...399..345E}
\begin{equation}
J_{i}=\left\{j\mid L_j\geqslant L_i,z_j\leqslant z_i^{\max}\right\},
\end{equation}
where $L_i$ is the $i$th FRB luminosity and $z_i^{\max}$ is the maximum redshift where the FRB with luminosity $L_i$ can be detected. We plot this region in Figure ~\ref{L-z} as a black rectangle. We define the number of FRBs contained in this region as $n_i$. Moreover, $N_i=n_i-1$ means taking the $i$th FRB out. This is why $c^-$ is in \cite{1971MNRAS.155...95L}. 

Similarly, we can define $J_i^{\prime}$  as
\begin{equation}
J_{i}^{\prime}=\left\{j\mid L_{j}\geqslant L_{i}^{\mathrm{lim}},\:z_{j}<z_{i}\right\}.
\end{equation}
In this equation, $z_i$ means the redshift of $i$th FRB and $L_i^\mathrm{lim}$ is the minimum observable luminosity at this redshift. This region is displayed as the red rectangle in Figure ~\ref{L-z}. Also we define the number of events included in this region as $M_i$. 

\begin{figure}
\centering
	\includegraphics[width=0.5\textwidth]{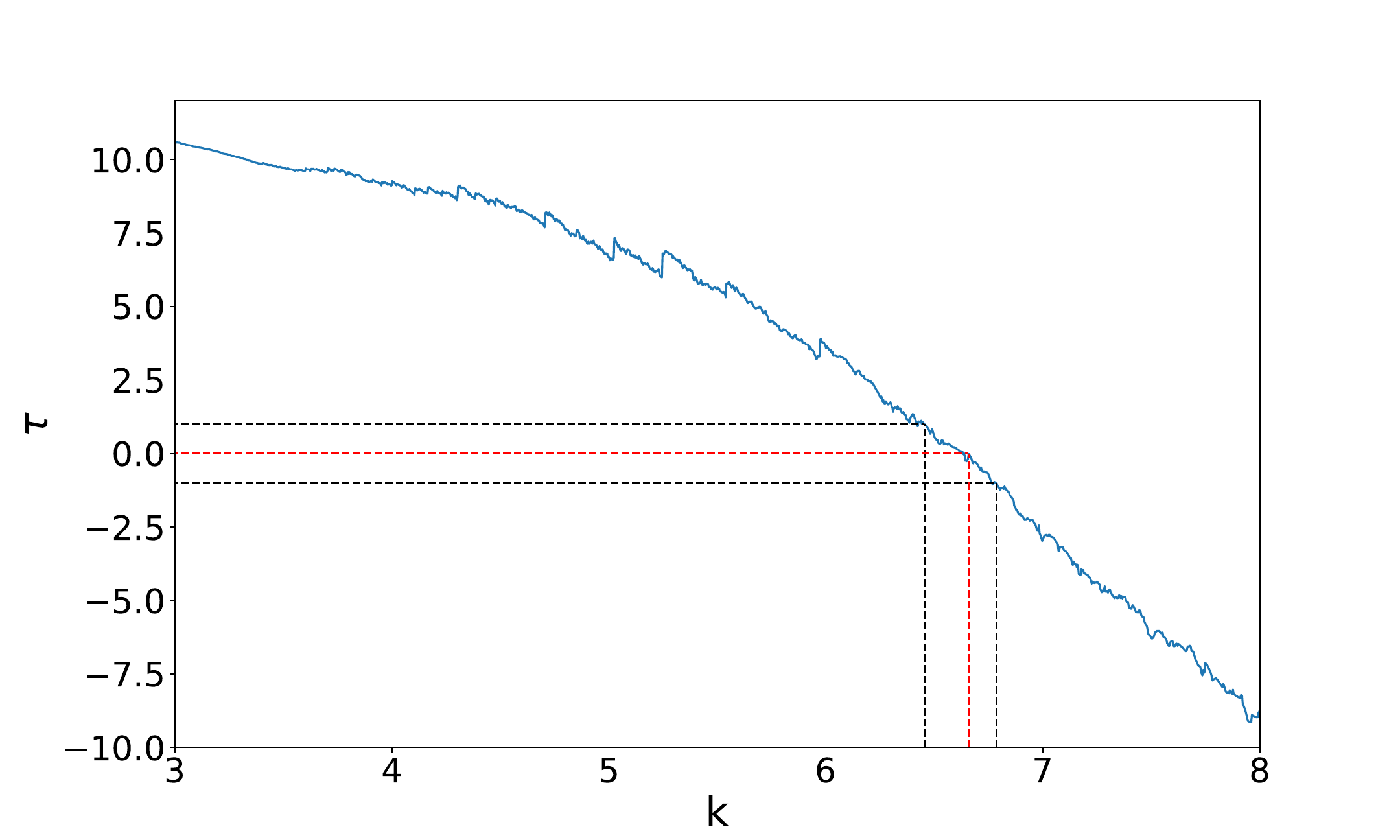}
    \caption{Value of test statistic $\tau$ as a function of $k$. The red dotted line is the best fit for $\tau=0$ and the black dotted lines represent $1\sigma$ errors. The optimal value is $k=6.66_{-0.21}^{+0.13}$ at $1\sigma$ confidence level.}
    \label{k-tao}
\end{figure}

We consider the $n_i$ FRBs contained in the black rectangle in Figure ~\ref{L-z} and define the number of events with redshifts equal to or less than $z_i$ as $R_i$. In addtion, we introduce the expected mean and the variance of $R_i$ as $E_i=\frac{n_i+1}2$ and $V_i= \frac{n_i^2-1}{12}$ respectively. If the luminosity and the redshift are independent, the $R_i$ should be uniformly distributed between 1 and $n_i$. The test statistic $\tau$ is \citep[]{1992ApJ...399..345E}
\begin{equation}
\tau=\sum_i\frac{(R_i-E_i)}{\sqrt{V_i}}.
\end{equation}
When the test statistic $\tau$ is zero, the luminosity and the redshift are independent. So we change the value of $k$ until $\tau$ is zero and get the form of $g(z)$. Figure ~\ref{k-tao} shows the test statistic $\tau$ as a function of $k$ and we find the best fit is $k=6.66_{-0.21}^{+0.13}$ within $l\sigma$ confidence level.
 
After removing the evolution's effect through $L_0=L/(1+z)^{6.66}$, the cumulative luminosity function $\psi(L_0)$ can be derived from the following equation with a non-parametric method \citep[]{1971MNRAS.155...95L,1992ApJ...399..345E}:
\begin{equation}
\psi(L_{0i})=\prod_{j<i}(1+\frac1{N_j}).
\end{equation}
In this equation, $j<i$ means the luminosity of $j$th FRB is larger than the $i$th FRB's. Similarly, we can obtain the cumulative redshift distribution $\phi(z)$ from
\begin{equation}
\phi(z_i)=\prod_{j<i}(1+\frac1{M_j}).
\end{equation}
In this equation, $j<i$ means the redshift of $j$th FRB is smaller than the $i$th FRB's.

\section{Luminosity function and formation rate}
\label{Luminosity function and formation rate of FRBs}
In this section, we will show results about the luminosity function and formation rate of FRBs with Lynden-Bell's $c^{-}$ method. 

\subsection{Luminosity function}
\begin{figure}
\centering
	\includegraphics[width=0.5\textwidth]{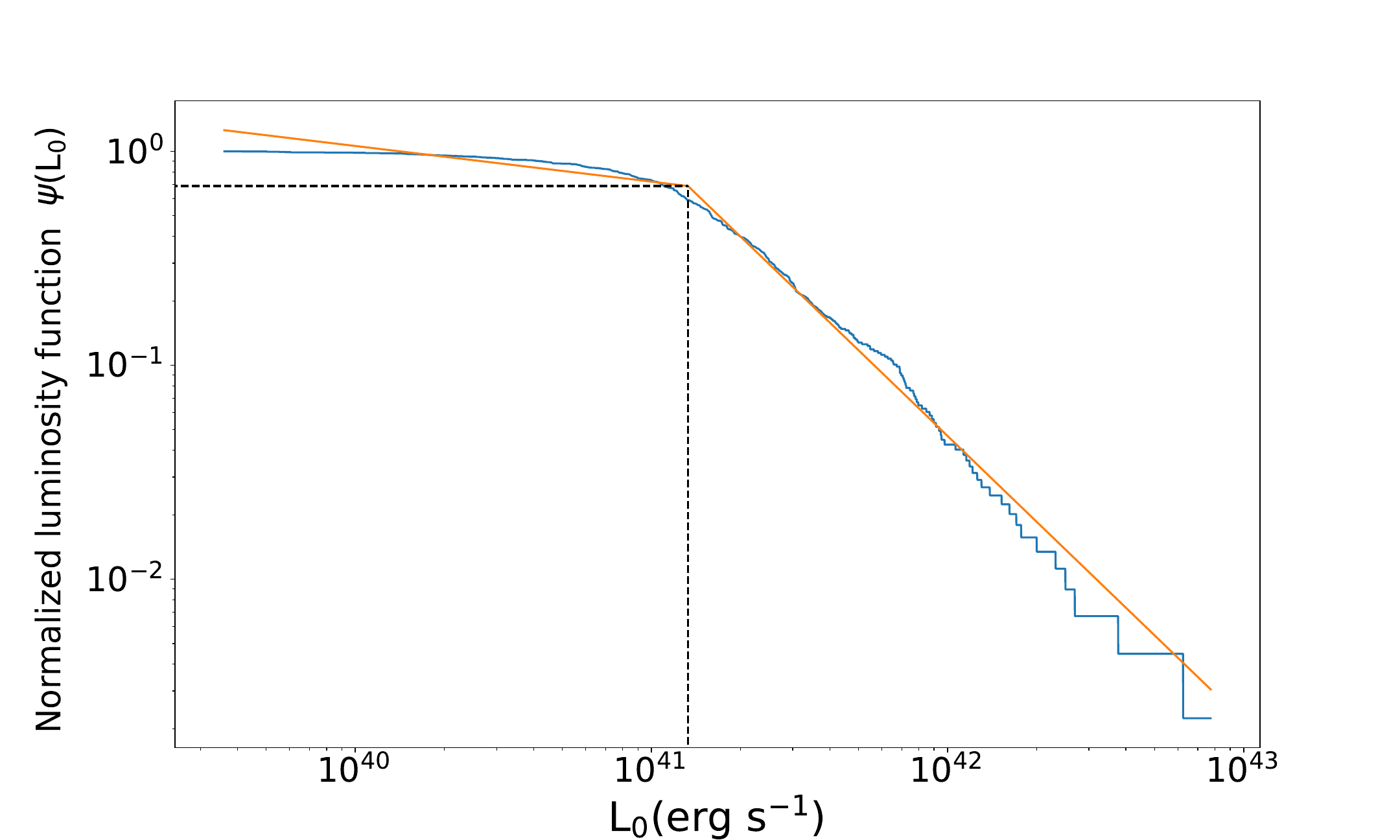}
    \caption{Cumulative luminosity function $\psi(L_0)$. The blue solid line represents observed data which is normalized to unity at the first point and the red line represents the best fit with a broken power-law model.}
    \label{Luminosity function.pdf}
\end{figure}

As discussed in Section \ref{Lynden-bell's}, we have already obtained the luminosity evolution $g(z)=(1+z)^{6.66}$. We get $L_{0}$ with $L_0=L/g(z)$ for Catalog 1, which is shown in Figure ~\ref{L0-z.pdf}. After having this new data set, the local luminosity function $\psi(L_0)$ can be obtained with Lynden Bell's $c^-$ method, which is displayed in Figure ~\ref{Luminosity function.pdf}. The luminosity function $\psi(L_0)$ can be fitted with a broken power law model. The best fit for dim and bright bursts is given by
\begin{equation}
\psi(L_0)\propto\begin{cases}L_0^{-0.17\pm0.01}&L_0<L_0^b\\L_0^{-1.33\pm0.01}&L_0>L_0^b\end{cases},
\label{luminosity}
\end{equation}
where $L_0^b=1.33\times10^{41}$ erg s$^{-1}$ is the break luminosity point. It is necessary to be noted that this is the luminosity function at $z=0$, because the luminosity evolution is removed. The luminosity function $\psi_{z}(L)$ at redshift $z$ should be $\psi_z(L)=\psi(L_0)(1+z)^{6.66}$. Similarly, the corresponding broken luminosity at $z$ is $L_z^b=L_0^b(1+z)^{6.66}$.

\begin{figure}
\centering
	\includegraphics[width=0.5\textwidth]{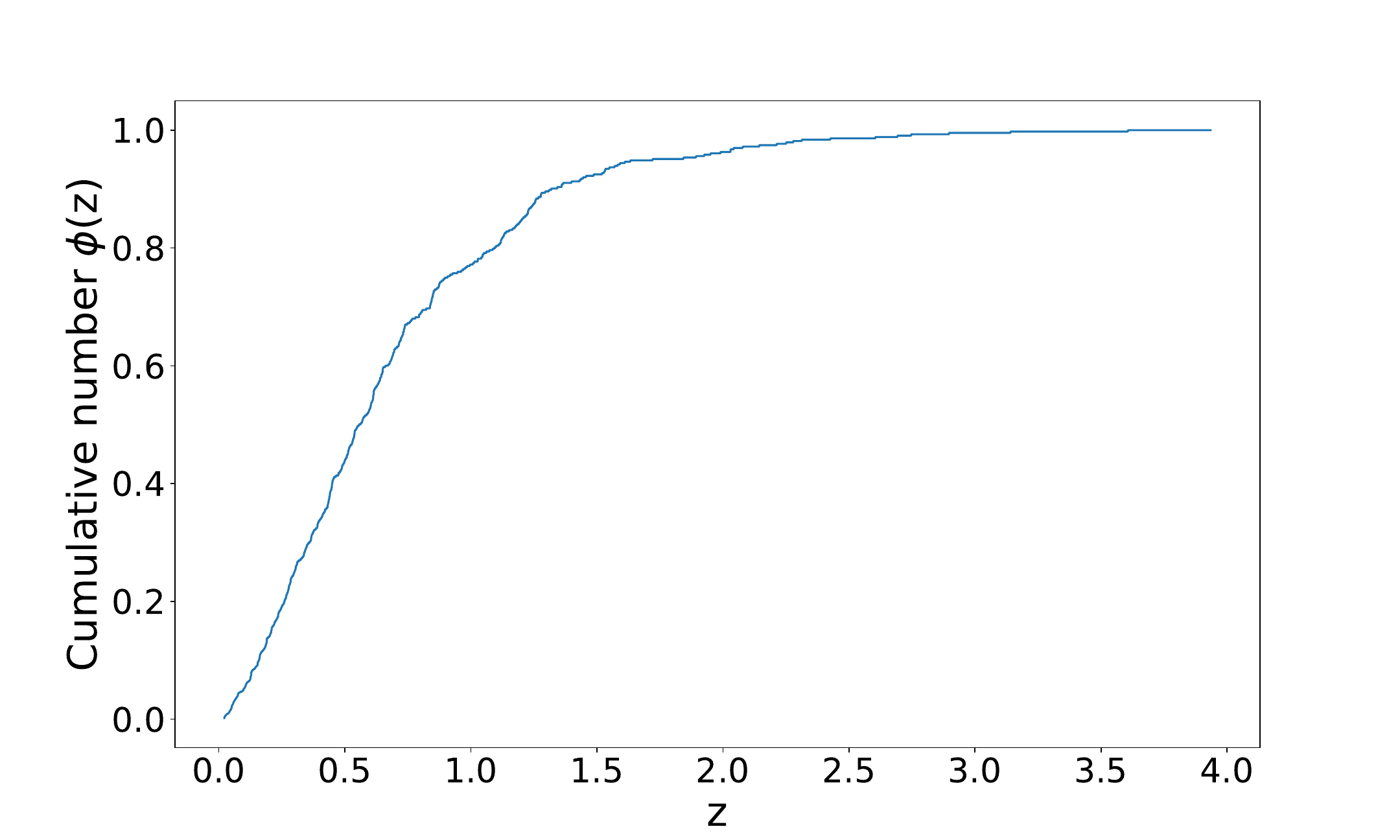}
    \caption{Normalized cumulative redshift distribution of FRBs, which is normalized to unity at the last point.}
    \label{Normalized cumulative redshift distribution.pdf}
\end{figure}

\begin{figure}
\centering
	\includegraphics[width=0.5\textwidth]{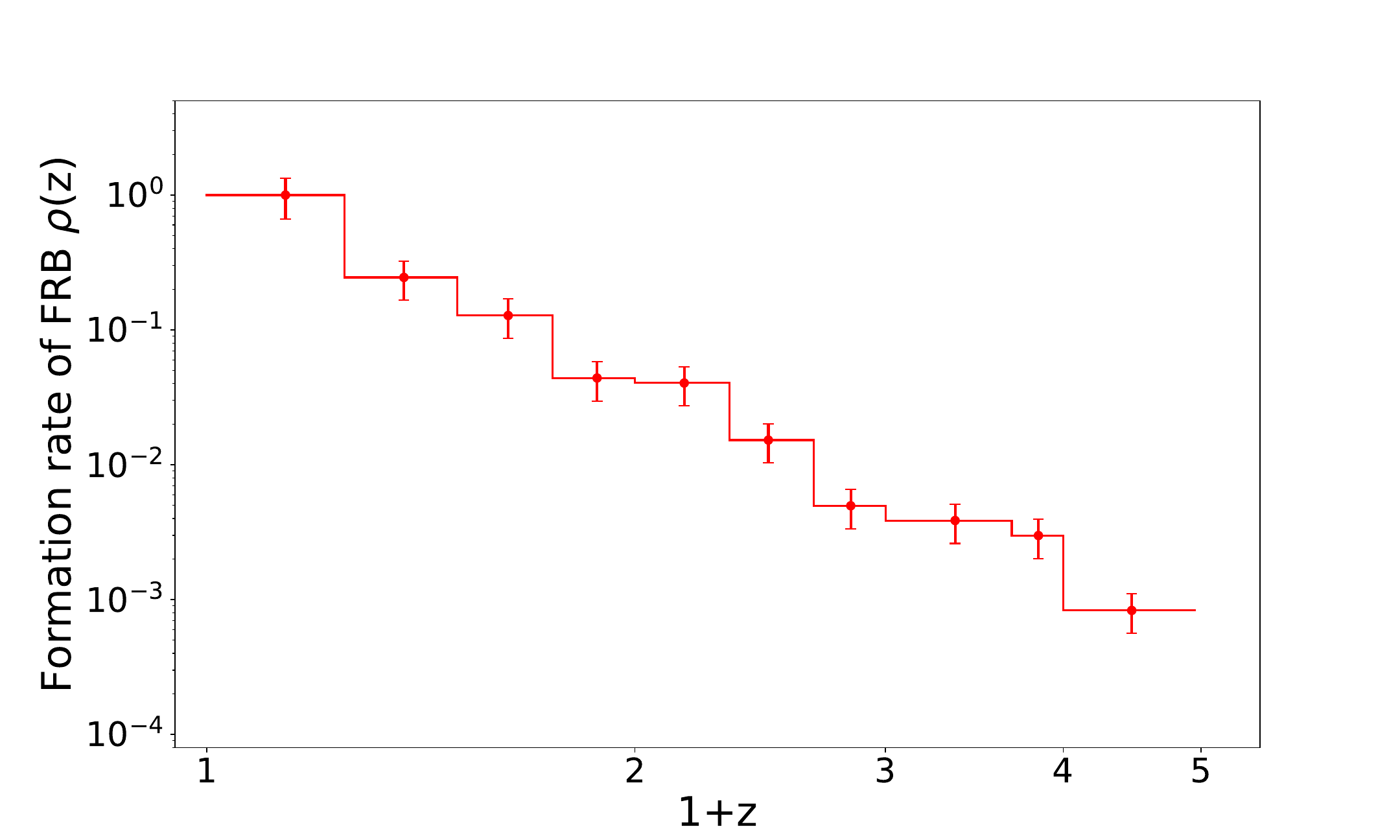}
    \caption{Comoving formation rate $\rho(z)$ of FRBs. It is normalized to unity at the first point and the $1\sigma$ error is also shown.}
    \label{Formation rate of FRBs.pdf}
\end{figure}

\subsection{Formation rate of FRBs}
Figure ~\ref{Normalized cumulative redshift distribution.pdf} shows the cumulative redshift distribution $\phi(z)$ of FRBs. Compared with GRBs, the formation rate of FRBs can be derived from
\begin{equation}
\rho(z)=\frac{d\phi(z)}{dz}(1+z)\Bigg(\frac{dV(z)}{dz}\Bigg)^{-1}.
\end{equation}
In this equation, the factor $(1+z)$ results from the cosmological time dilation and the differential comoving volume $dV(z)/dz$ can be expressed as
\begin{equation}
\begin{aligned}
\frac{dV(z)}{dz}&=4\pi\Bigg(\frac{c}{H_0}\Bigg)^3\Bigg(\int_0^z\frac{dz}{\sqrt{1-\Omega_{\mathrm{m}}+\Omega_{\mathrm{m}}(1+z)^3}}\Bigg)^2  \\
&\times\frac1{\sqrt{1-\Omega_{\mathrm{m}}+\Omega_{\mathrm{m}}(1+z)^{3}}}.
\end{aligned}
\end{equation}
After calculating the cumulative redshift distribution with the above formulae, we get the formation rate of FRBs shown in Figure ~\ref{Formation rate of FRBs.pdf}. The best fitting power-law model for $\rho(z)$ is
\begin{equation}
\rho(z)\propto(1+z)^{-4.9\pm0.3},
\label{redshift}
\end{equation}
with $1\sigma$ confidence level. From this equation, we can derive the formation rate of FRBs in the local universe $\rho(0)$ is $1.13\times10^4\ \mathrm{Gpc}^{-3} \ \mathrm{yr}^{-1}$. The formation rate $\rho(z)$ decreases monotonically with redshift. Instead, \citet{2019JHEAp..23....1D} found that FRB rate is consistent with SFR. The possible reason is that the small sample is used in their work (17 and 21 FRBs), which can not give a reliable statistical result. 

\begin{figure*}
\centering
	\includegraphics[width=\textwidth]{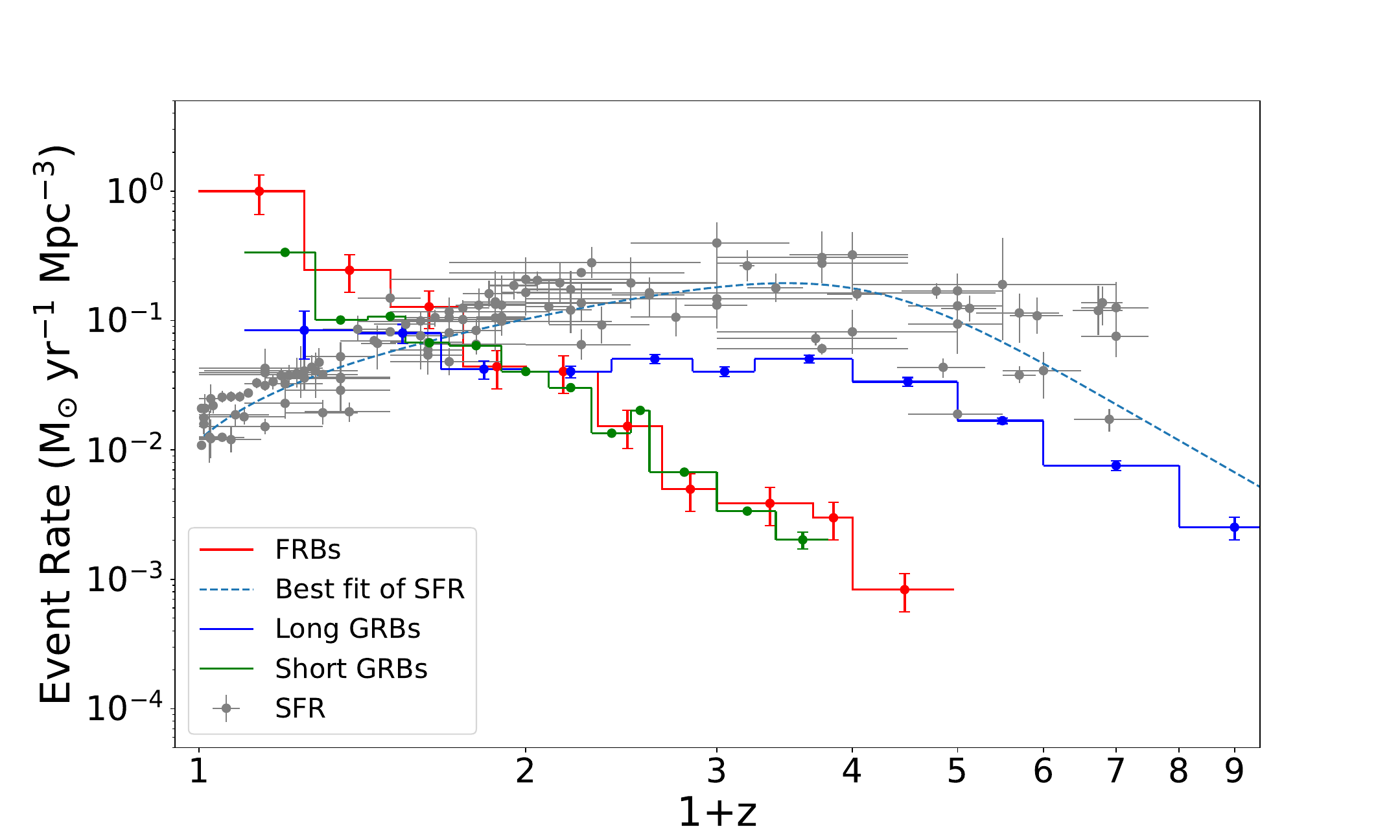}
    \caption{Comparison between the formation rate of FRBs and other events. The red line represents formation rate of FRBs. The green and blue lines represent the rates of short GRBs \citep[]{2018ApJ...852....1Z} and long GRBs \citep[]{2015ApJS..218...13Y}, respectively. Gray dots and the blue dashed line correspond to the observed SFR and the best fit \citep[]{2006ApJ...651..142H}.}
    \label{Event Rate comparison.pdf}
\end{figure*}

\begin{figure}
\centering
	\includegraphics[width=0.5\textwidth]{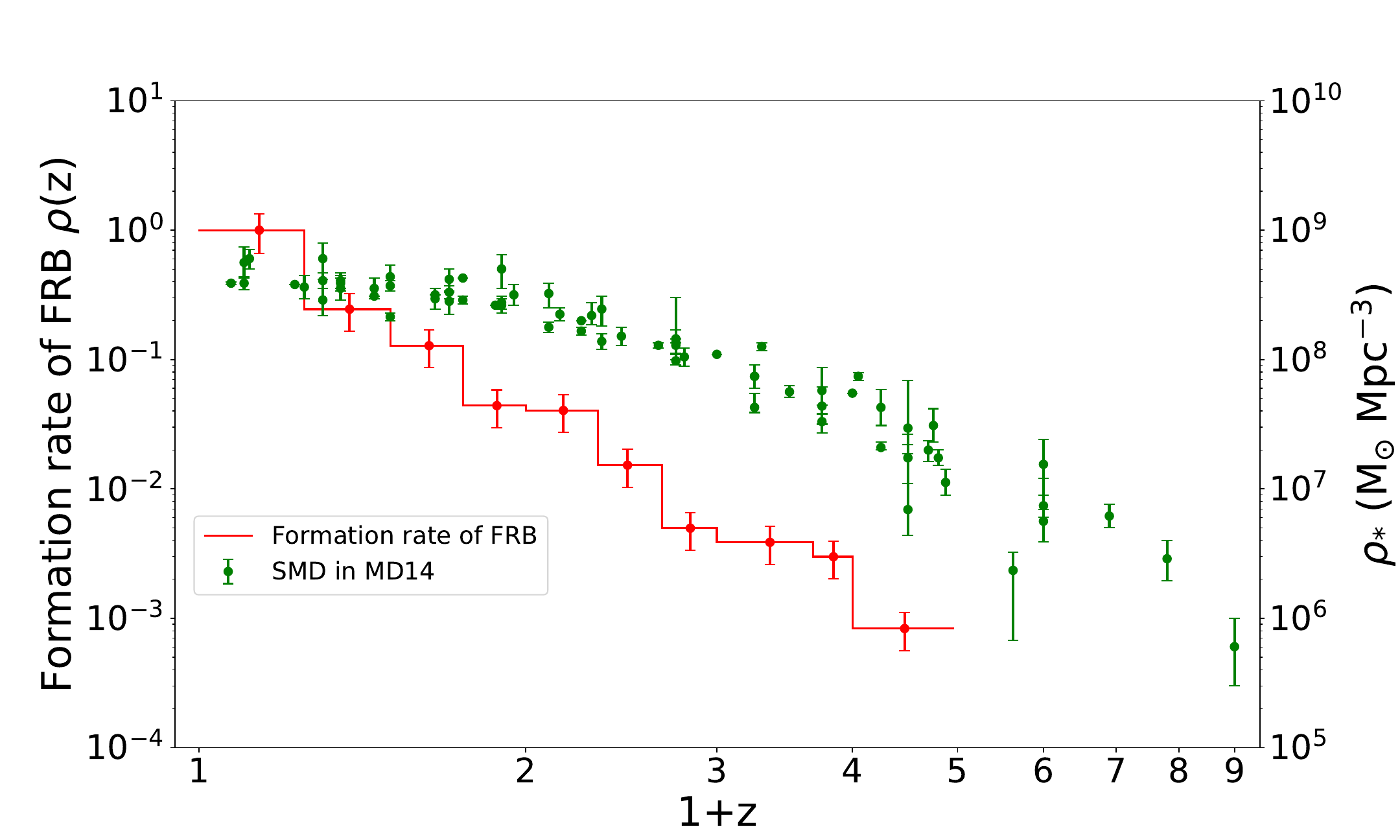}
    \caption{Comparison between FRB rate and the observed SMD. The red line represents the formation rate of FRBs, and the green dots represents the observed SMD \citep[]{2014ARA&A..52..415M}. All error bars are $1\sigma$ confidence level.}
    \label{Event Rate comparison2.pdf}
\end{figure}

The relation between the rate of FRBs and the SFR is attractive. If their origins are related to the deaths of massive stars, the FRB rate should track the SFR. So, we compare our result with the observed SFR \citep[]{2006ApJ...651..142H} in Figure ~\ref{Event Rate comparison.pdf}. It is obvious that the FRB rate deviates from SFR. We also compare our result with long GRBs \citep[]{2015ApJS..218...13Y} and short GRBs \citep[]{2018ApJ...852....1Z}. The redshift-dependence of FRB rate is similar as short GRBs. But their local formation rates are significantly different, which indicates compact binary mergers may only occupy a portion in the origin of FRBs.

In Figure ~\ref{Event Rate comparison2.pdf}, we compare the formation rate of FRBs and the stellar mass density (SMD) \citep[]{2014ARA&A..52..415M}. The formation rate of FRBs decreases more quickly with redshift than SMD. But they show similar decreasing trend at $z>1.0$. Our comparison suggests that non-repeating FRBs may be mainly related to old populations rather than young populations which can be tracked by the SFR. This indicates old neutron stars or black holes are more likely to be progenitors.

\section{Testing with Monte Carlo simulation}
\label{Testing with Monte Carlo simulation}
\begin{figure*}
\centering
\subfigure[]{
\includegraphics[width=0.45\textwidth]{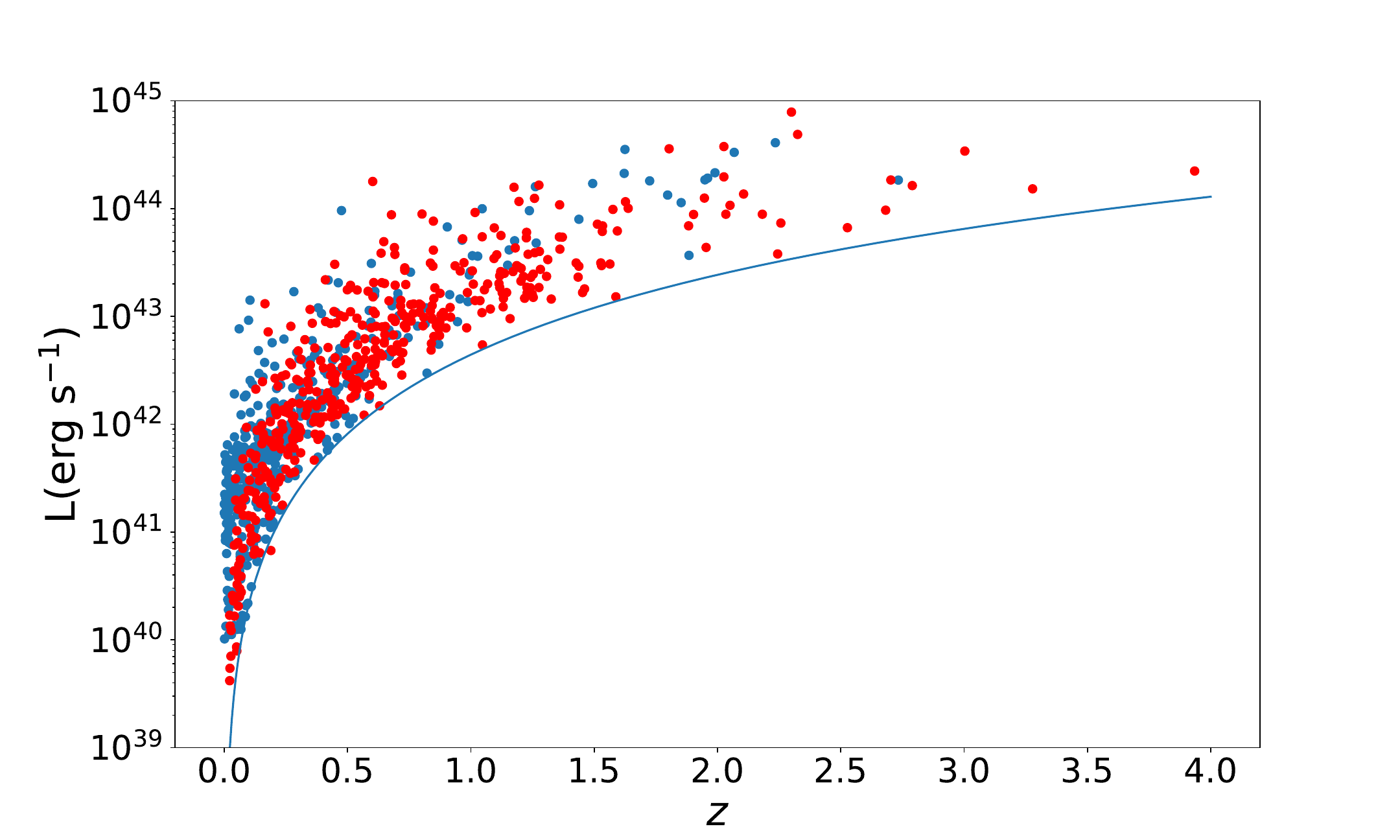}
}
\quad
\subfigure[]{
\includegraphics[width=0.45\textwidth]{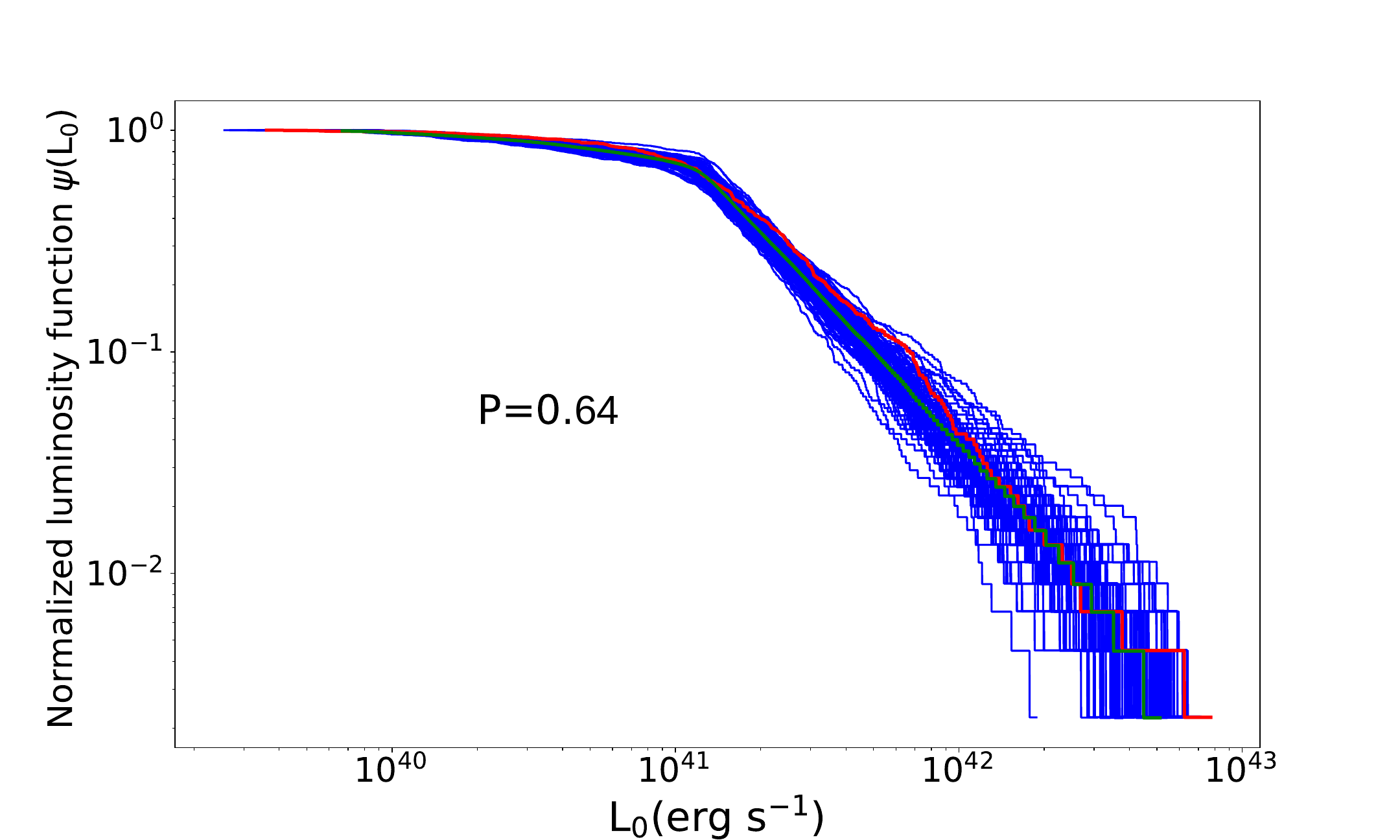}
}
\quad
\subfigure[]{
\includegraphics[width=0.5\textwidth]{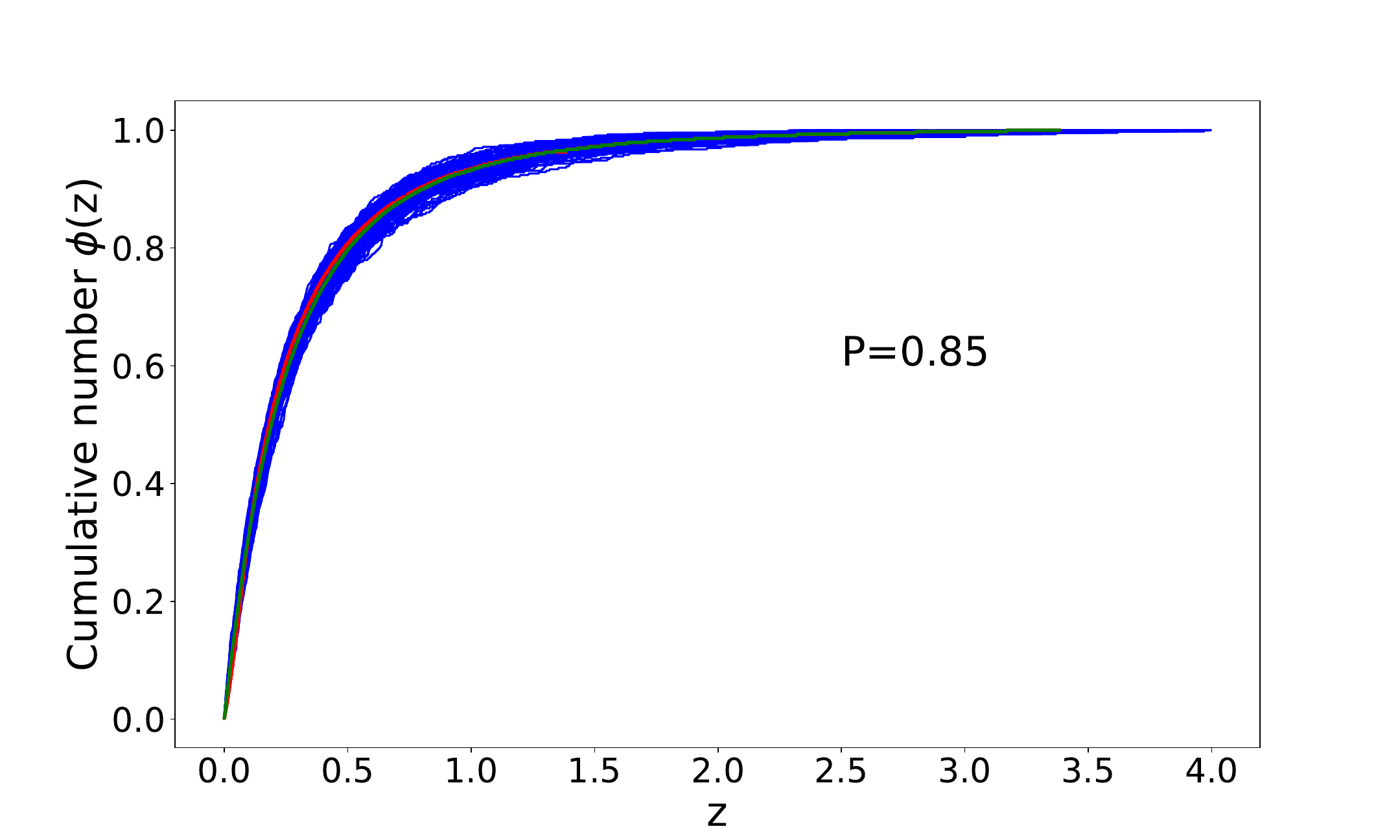}
}
\quad

\caption{Comparison of the simulated data (blue) and the observed data (red). Panel (a) shows the luminosity–redshift distribution. We choose one sample from 100 simulated samples randomly. The red points and blue points represent the observed data and simulated data respectively.
For panels (b) and (c), blue lines are the results of 100 samples. The green lines are the mean distributions of those simulated samples. The probabilities of Kolmogorov–Smirnov tests between the mean simulated data and the observed data are also given.}
\label{Monte carlo simulation}
\end{figure*}

In this part, we will test our results with Monte Carlo simulation. First, we generate a set of FRB data with luminosity and redshift $(L_{0},z)$ by Monte Carlo method, whose two parts satisfy the distribution described by equations (\ref{luminosity}) and (\ref{redshift}). As we mentioned above, the luminosity function of equation (\ref{luminosity}) is at $z=0$. So, we should transfer $L_0$ to $L= L_0( 1+ z)^k$, where we choose $k= 6.66$ obtained before. Then, we can get a set of pseudo points $(L,z)$, which is similar to the observed data shown in Figure \ref{Monte carlo simulation}. In the simulation, we generate 100 pseudo samples totally. Each sample contains 439 FRBs, which is same as the selected CHIME/FRB sample. Then we perform the same analysis with Lynden-Bell's $c^{-}$ method to process these simulated samples to obtain the luminosity function and formation rate of FRBs. In the end, we compare the simulated data with the observed data using the Kolmogorov–Smirnov test. 

In Figure ~\ref{Monte carlo simulation}, we display the results of Monte Carlo simulation. In the panel (a), we randomly choose one sample of FRBs from 100 samples to compare with selected CHIME/FRB data and show the comparing result of the luminosity–redshift distribution. The red points and blue points represent the observed data and simulated data, respectively. The blue solid line is the flux limit of 0.2 Jy. They have the same distribution and the distribution of the simulated data is located in the area of the observed data. In panels (b) and (c), the blue solid lines are simulated results of the cumulative luminosity function and the cumulative redshift distribution respectively. Kolmogorov–Smirnov test results are 0.64 and 0.85, between the mean distribution of simulated data (the green line) and the distribution of observed data (the red line). So, the derived cumulative luminosity function and formation rate are reliable.

\section{Conclusions and discussion}
\label{Conclusions and discussions}
In this paper, we use Lynden-Bell's $c^-$ method to study the luminosity function and formation rate of the sample without any assumptions. The luminosity function of FRBs can be well fitted with a broken power-law model, $\psi(L_0)\propto L_0^{-0.17\pm0.01}$ for dim FRBs and $\psi(L_0)\propto L_0^{-1.33\pm0.01}$ for bright FRBs, where the break point is $1.33\times10^{41}$\ erg s$^{-1}$. The formation rate decreases monotonically as $\rho(z)\propto(1+z)^{-4.9\pm0.3}$ with the local formation rate $1.13\times10^4\ \mathrm{Gpc}^{-3} \mathrm{yr}^{-1}$. 

FRB redshift evolution can provide us with some evidence to constrain its progenitors statistically. If young stellar populations involve FRB events, the FRB rate should follow star formation history increasing towards higher redshifts. If old populations such as old neutron stars and black holes dominate the generation of FRBs, FRB rate is expected to decrease towards higher redshifts. In this work, Lynden-Bell's $c^-$ method can break the degeneracy between the luminosity function and formation rate, and the first CHIME/FRB catalogue provide us with a larger and homogeneous FRB sample. After comparing the FRB rate with the SFR, the rate of long GRBs and short GRBs, and SMD, we find that the old populations including neutron stars and black holes, are closely related to the origins of FRBs. Our conclusion is basically consistent with previous research on Catalog 1 \citep{2022MNRAS.511.1961H, 2022ApJ...924L..14Z}, namely that FRBs do not track star formation of the universe and can be better described by delayed models. 

There are some models dealing with how to make FRBs with old populations. The first possibility is that FRBs are generated by reactivation of old neutron stars \citep{Beniamini2020,Wadiasingh2020,Lan2023} or binary neutron star (white dwarf) mergers \citep{Margalit2019,Wang2020,Lu2022,Kremer2023}. However, the fractional contribution to the total FRB rate by compact binary mergers is a few percent \citep{ZhangG2020,Lu2022}.
The second possibility is that FRBs are produced by black holes \citep{Katz2020,Sridhar2021}. 
Third, FRBs are powered by old neutron stars interacting with other objects \citep{Dai2016,Zhang2017,Wang2022}. 
Further study on these models may be required.


\section*{acknowledgments}

We thank Guo-Qiang Zhang for helpful discussions. This work was supported by the National Natural Science Foundation of China (grant No. 12273009), and the National SKA Program of China (grant No. 2022SKA0130100). We acknowledge the use of CHIME/FRB Public Database, provided at https://www.chime-frb.ca/ by the CHIME/FRB Collaboration.

\bibliography{sample631}{}
\bibliographystyle{aasjournal}



\end{document}